\documentclass{article}
\usepackage{spconf,amsmath,graphicx,hyperref,bbm}

\title{Towards Robust Dysarthric Speech Recognition: LLM-Agent Post-ASR Correction Beyond WER}
\name{%
  Xiuwen Zheng$^{1}$,
  Sixun Dong$^{2}$,
  Bornali Phukon$^{1}$,
  Mark Hasegawa-Johnson$^{1}$,
  Chang D. Yoo$^{3}$
}
\address{%
  $^{1}$Dept. of ECE, University of Illinois Urbana-Champaign, IL, USA \\
  $^{2}$Independent Researcher, AZ, USA \\
  $^{3}$Dept. of EE, Korea Advanced Institute of Science \& Technology, KR \\
  \{xiuwenz2, bornalip, jhasegaw\}@illinois.edu, sixundong.ai@gmail.com, cd\_yoo@kaist.ac.kr
}

\usepackage[T1]{fontenc}
\usepackage{xcolor}
\usepackage{booktabs}
\usepackage{multirow}
\usepackage{algorithm}
\usepackage{algpseudocode}
\usepackage{tabularx}
\usepackage{enumitem}
\usepackage{makecell}
\usepackage{multirow}
\usepackage{dblfloatfix}
\usepackage{placeins}
\usepackage{amsmath,amssymb}
\usepackage{pifont}
\hypersetup{pdfborder={0 0 0}}

\newcommand\graycross{\textcolor[rgb]{ .502,  .502,  .502}{\ding{55}}}

\begin{document}
%\ninept
%
\maketitle
\begin{abstract}
    While Automatic Speech Recognition (ASR) is typically benchmarked by word error rate (WER), real-world applications ultimately hinge on semantic fidelity. This mismatch is particularly problematic for dysarthric speech, where articulatory imprecision and disfluencies can cause severe semantic distortions. To bridge this gap, we introduce a Large Language Model(LLM)-based agent for post-ASR correction: a \textbf{Judge–Editor} over the top-\(k\) ASR hypotheses that keeps high-confidence spans, rewrites uncertain segments, and operates in both zero-shot and fine-tuned modes. In parallel, we release \textbf{SAP-Hypo5}\footnote{\url{https://github.com/xiuwenz2/SAP-Hypo5}}\footnote{\url{https://huggingface.co/datasets/xiuwenz2/SAP-Hypo5}}, the largest benchmark for dysarthric speech correction, to enable reproducibility and future exploration. Under multi-perspective evaluation, our agent achieves a 14.51\% WER reduction alongside substantial semantic gains, including a \(+7.59\,\mathrm{pp}\) improvement in MENLI and \(+7.66\,\mathrm{pp}\) in Slot Micro F1 on challenging samples. Our analysis further reveals that WER is highly sensitive to domain shift, whereas semantic metrics correlate more closely with downstream task performance.

\end{abstract}

\begin{keywords}
Post-ASR Correction, Dysarthric Speech, LLM Agent, Semantic Fidelity, Robust Speech Recognition
\end{keywords}
%
% \section{Introduction}

\section{Introduction}  \label{sec:Intro}
Large Language Model(LLM)-based post-ASR correction has recently shown promise on standard English~\cite{ma2023can, chen2023hyporadise, yang2024large}, yet most studies optimize and report only word error rate (WER), the dominant metric for automatic speech recognition (ASR). However, many real-world applications-captioning, note-taking, and downstream spoken language understanding (SLU), ultimately depend on the preservation of meaning. In dysarthric speech recognition and assistive communication, reduced articulatory precision, frequent disfluencies and atypical prosody often make ASR generate plausible words that miss the speaker’s intended meaning. The Speech Accessibility Project Challenge (SAPC~\cite{zheng25_interspeech}) scored systems using a measure called SemScore~\cite{phukon2025aligning} that was a combination of phonetic, lexical semantic, and natural language inference. LLM post-ASR correction sometimes improved SemScore at the expense of WER~\cite{la2025exploring}, but averaging over the test set masked such differences~\cite{zheng25_interspeech}, suggesting that spoken language understanding may require greater measurement precision.

\begin{figure}
    \centering
    \includegraphics[width=0.98\linewidth]{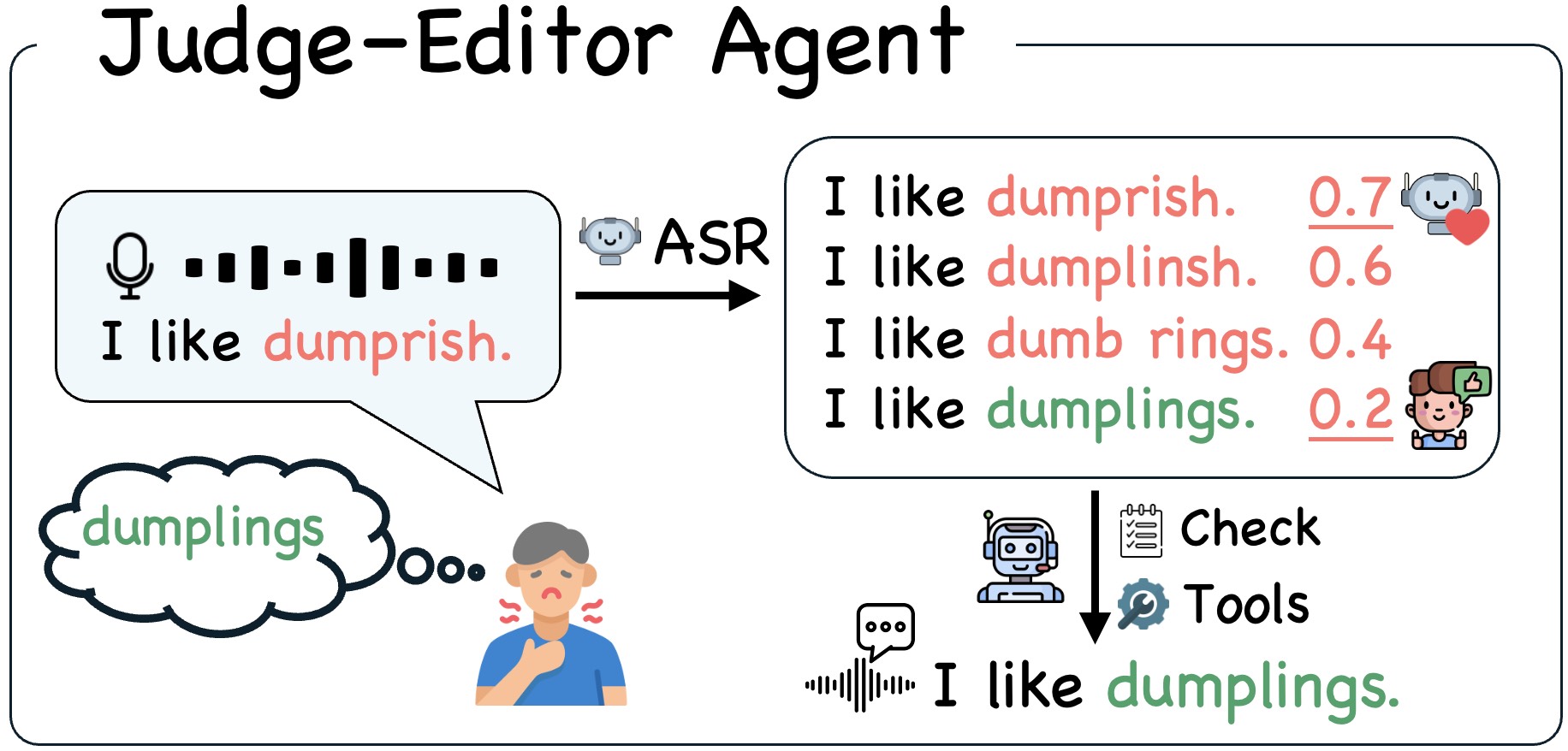}
    \caption{\textbf{Judge–Editor Agent (JEA)}. Given the ASR multiple hypotheses, JEA first judges span-level uncertainty and cross-hypothesis consistency, then editing and fusion spans to synthesize a single intent-preserving transcript.
    }
    \label{fig:framework}
\end{figure}
To enhance semantic fidelity of dysarthric speech recognition, we reformulate post-ASR correction as an \emph{agentic decision problem}: given the ASR top-$k$ hypotheses, an LLM acts as a \textbf{Judge-Editor} that retains high-confidence spans while selectively rewriting or fusing uncertain segments to better capture the intent of the speaker. The approach leaves the acoustic model untouched and remains training-light, operating in either zero-shot or lightly fine-tuned settings. In parallel, we introduce \textbf{SAP-Hypo5}, a new benchmark for dysarthric post-ASR correction derived from the Speech Accessibility Project (SAP). The dataset comprises 35k utterances, each paired with a gold reference transcript and the ASR top-$5$ unique hypotheses. Furthermore, we design a fine-grained, multi-metric evaluation protocol that integrates semantic similarity and downstream SLU metrics, enabling more sensitive and informative assessment of system performance.

Our framework demonstrates strong performance on SAP-Hypo5, confirming that post-ASR correction with an LLM-based Judge–Editor is training-light but effective. Beyond WER, semantic metrics and downstream task-oriented measures provide extra robustness under domain shift. We summarize our contributions as follows:
\begin{itemize}[nosep,leftmargin=*]
\item \textbf{Method.} We propose a simple yet effective \textbf{ASR$\rightarrow$LLM-agent} system that reads top-$k$ candidates and performs span-level keep, rewrite, or fuse decisions. It is training-light and requires no changes to the acoustic model.
\item \textbf{Benchmark.} We release \textbf{SAP-Hypo5}, a 35k-utterance dysarthria benchmark with standardized splits and normalization, where each utterance is paired with its reference transcript and the top-5 \emph{unique} ASR hypotheses.
\item \textbf{Evaluation.} We introduce a fine-grained, multi-metric evaluation setup including {WER}$\downarrow$, {Semantic} metrics (Q-Emb$\uparrow$, BERTScore$\uparrow$, MENLI$\uparrow$), and downstream {SLU} metrics (Intent Accuracy$\uparrow$, Slot Micro F1$\uparrow$), providing a more comprehensive measure of system performance.
\end{itemize}

\section{Dataset}
\label{sec:Dataset}

Prior work on LLM-based ASR correction (e.g., HyPoradise~\cite{chen2023hyporadise}) establishes strong baselines on standard English, but there is still no open dataset for dysarthria. 
We therefore introduce \textbf{SAP-Hypo5}, which pairs each SAP~\cite{hasegawa2024community} utterance with its reference transcript and the top-$5$ hypotheses from an open-source ASR fine-tuned on SAP (Challenge release, PD only)\footnote{\url{https://huggingface.co/xiuwenz2/whisper-large-v2-ft-SAP-0430-pd-merged-ct2}}. By capturing both cross-speaker variability and the atypical acoustic–phonetic patterns of dysarthria, SAP-Hypo5 provides a reproducible testbed for investigating semantic preservation under challenging dysarthric conditions.

Hypotheses are generated using beam search. The top-1 hypothesis from this process serves as the ASR baseline for our subsequent experiments. From 50 beam candidates, we retain the five highest-scoring unique hypotheses; if fewer than five are distinct, we follow~\cite{chen2023hyporadise} and fill the remainder by random sampling. 
We use the SAP train/dev splits (SAP test is unavailable) and further divide dev into our dev/test. Since SAP is speaker-independent, with no speaker overlap between train and dev, SAP-Hypo5 naturally inherits this property.
To align with~\cite{chen2023hyporadise}, we keep utterances of length 4–32 words, deduplicate, and normalize transcripts: preserve quotation marks, remove other punctuation, split abbreviations (e.g., ``T V''), and lowercase. 
Any utterance whose transcript duplicates training text is excluded from the dev and test sets. As a result, the final splits consist of 31123, 845, and 2647 utterances for training, development and test, respectively.

\section{Methodology}
\label{sec:Methodology}

We formulate post-ASR correction as an \emph{agentic decision problem}. 
Given an utterance, the ASR system produces top-$k$ hypotheses 
$\{h_1, h_2, \dots, h_k\}$ as candidate transcripts. 
Our goal is to generate a corrected output $\hat{y}$ that maximizes semantic 
fidelity to the speaker’s intended transcript $y_{\text{ref}}$. 
To this end, we present the \textbf{ASR$\rightarrow$LLM-agent} framework, 
where an LLM serves as a \textbf{Judge-Editor}: it inspects disagreements across hypotheses, 
preserves reliable spans, and selectively rewrites or fuses uncertain segments. 
The formulation is \emph{model-agnostic}, requiring no changes to the acoustic model.

\vspace*{-10pt}
\subsection{Implementation}
\vspace*{-6pt}
We consider two deployment modes for the \textbf{Judge-Editor Agent (JEA)}, zero-shot prompting and lightweight task-adaptive fine-tuning.

In the zero-shot setting, the LLM is prompted to serve as a \textbf{Judge-Editor} over the ASR top-$k$ hypotheses without task-specific training. The prompt enforces a simple policy: preserve spans that show high cross-hypothesis agreement; rewrite or fuse only uncertain segments; and avoid hallucinating content, especially named entities and numbers. In practice, we use Qwen2-7B-Instruct\cite{yang2024qwen2} with frozen weights with an iteratively refined instruction template.

To better internalize dysarthria-specific patterns, we perform lightweight self-supervised fine-tuning on SAP-Hypo5, following the general configuration of the GenSEC Challenge at IEEE SLT 2024~\cite{yang2024large}. We fine-tuned multiple agents, including Qwen2-7B\cite{yang2024qwen2}, Qwen3-8B\cite{yang2025qwen3}, Llama-2-7B\cite{touvron2023llama}, and Llama-3-8B\cite{dubey2024llama}, and additionally evaluate the released baseline model\footnote{\url{https://huggingface.co/GenSEC-LLM/SLT-Task1-Llama2-7b-HyPo-baseline}} from ~\cite{yang2024large} (hereafter referred to as Llama-2-7B-H), fine-tuned on HyPoradise~\cite{chen2023hyporadise}, a large benchmark built on standard speech corpora. Fine-tuning is performed under an agentic-instructional template, which enforces the \textbf{Judge-Editor} role, using LoRA\cite{hu2022lora} to update $<0.25\%$ of parameters under \texttt{int8} quantization. Each training instance is a pair $\big(\{h_1,\dots,h_k\}, y_{\text{ref}}\big)$, where the input is the ranked and de-duplicated ASR hypotheses and the target is the human reference.
Rather than computing loss over the entire input–output sequence, the objective is applied only to output tokens. This accelerates convergence and allows training to finish within three epochs on a single A100 GPU ($\approx$ 8 hours). During inference, deterministic greedy decoding is adopted for post-correction to reduce hallucinations and ensure reproducibility, and the computational overhead is nearly identical for zero-shot and LoRA-adapted agents.

To mitigate ASR hallucinations that inflate edit distance and distort evaluation, we employ Repeated Phrase Truncation (Alg.~\ref{alg:postprocess}). The algorithm identifies contiguous repetitions and selects the phrase maximizing repetitive coverage (the product of repetition count and phrase length). The agent output $\hat{y}$ is then truncated after the first occurrence of this phrase, effectively pruning hallucinated loops to ensure a reliable performance assessment.

\begin{algorithm}[t]
\caption{Repeated Phrase Truncation}
\label{alg:postprocess}
\begin{algorithmic}[1]
\Function{TruncateRepeatedPhrase}{$\hat{y}, m, M$}
    
    \State $\mathcal{U} \gets \{ u \subseteq \hat{y} : u \text{ is a contiguous phrase}, m \le |u| \le M, \textsc{Repeat}(\hat{y}, u) \ge 2 \}$ 
    \Comment{$\textsc{Repeat}$: max consecutive occurrences}
    
    \If{$\mathcal{U} = \emptyset$}
        \Return $\hat{y}$
    \EndIf

    \State $u^* \gets \arg\max_{u \in \mathcal{U}} (\textsc{Repeat}(\hat{y},u) \cdot |u|, -i_u, -|u|)$
    \Comment{$i_u$: first occurrence index; max coverage, then earliest, then shortest}

    \State \Return $\hat{y}[1 : i_{u^*} + |u^*|]$
    \Comment{truncate repeated phrase}

\EndFunction
\end{algorithmic}
\end{algorithm}

\vspace*{-10pt}
\subsection{Evaluation}
\vspace*{-6pt}

Word Error Rate (WER) is a common metric for assessing ASR performance. It quantifies transcription accuracy as the normalized edit distance between the ASR hypothesis and the ground-truth reference.

Semantic Score evaluates ASR outputs in terms of how well they preserve the intended meaning and contextual coherence at the utterance level. In this work, we adopt three semantic evaluation metrics: Q-Emb, BERTScore F1, and MENLI. Q-Emb provides a holistic measure of sentence-level semantic preservation by extracting embeddings from the Qwen3-Embedding-8B~\footnote{\url{https://huggingface.co/Qwen/Qwen3-Embedding-8B}} model and computing similarity scores between hypothesis and reference sentences. In contrast, BERTScore F1~\cite{zhang2019bertscore} leverages token-level contextualized embeddings, enabling a more fine-grained assessment of semantic alignment. Complementing these embedding-based approaches, MENLI~\cite{chen2023menli} is a logical entailment score, which adopts a natural language inference (NLI) framework to evaluate whether the logical content of the reference is entailed by the ASR hypothesis.

Task-oriented metrics assess transcript quality by their downstream utility for Spoken Language Understanding (SLU). Following \cite{zailan2023state}, we report two widely used measures: Intent Accuracy (Intent Acc.) and Slot Micro\mbox{-}F1 (Slot F1). Intent accuracy is the proportion of utterances for which the system correctly predicts the user’s intended action, providing a high-level indicator of intent recognition reliability. Complementarily, Slot Micro\mbox{-}F1 offers a fine-grained view of entity extraction by micro\mbox{-}averaging the harmonic mean of precision and recall across all slot types. In this work, intents and slots are extracted using the intent model and the slot model released by~\cite{kubis2023back}, respectively; both models are XLM\mbox{-}RoBERTa variants fine\mbox{-}tuned on MASSIVE~\cite{fitzgerald2022massive}. Their outputs on the reference transcripts serve as \emph{pseudo\mbox{-}gold} SLU labels. We then compute intent accuracy and slot F1 as follows:

\textbf{Intent accuracy.}
Let $z_i$ be the gold intent for utterance $i$ and $\hat{z}_i$ the predicted intent. Intent accuracy is

\begin{equation}\label{eq:intent-acc}
\mathrm{Acc}_{\text{intent}}
= \frac{1}{M}\sum_{i=1}^{M}\mathbbm{1}\!\left[\hat{z}_i = z_i\right].
\end{equation}

\textbf{Slot micro\mbox{-}F1.}
To avoid sensitivity to reference–hypothesis misalignment, slots are evaluated at the value level. Each slot is mapped to a pair $(t,\mathrm{norm}(v))$, where $t$, $v$ are the slot type and value, respectively; $\mathrm{norm}(\cdot)$ performs text normalization. For each utterance, we form multisets of such pairs from the reference and the hypothesis; micro\mbox{-}F1 is then computed from corpus\mbox{-}wide $\mathrm{TP}$, $\mathrm{FP}$, and $\mathrm{FN}$ derived by multiset intersection/difference.

\vspace*{-6pt}
\section{Results and Discussion}
\label{sec:Results and Discussion}
\begin{table}
  \centering
  \caption{SAP-Hypo5 test set ($N=2,647$) results using different Judge-Editor Agents, reported in WER (\%). The first row presents the ASR baseline (top-1 hypothesis). Results are partitioned into NoErr ($N=1,080$) and Err ($N=1,567$) subgroups based on whether the top-1 ASR hypothesis is error-free. All LLMs employ their 7B–8B scale variants to provide a fair comparison.}
  \label{table:results_by_wer}
  \renewcommand{\arraystretch}{1.0}

  \begin{tabularx}{1.0\linewidth}{p{1.8cm}|p{2.2cm}||*{3}{>{\centering\arraybackslash}X}}
    \toprule
    \multirow{2}{*}{\textbf{Method}} & \multirow{2}{*}{\textbf{Agent}}
      & \multicolumn{3}{c}{\textbf{WER (\%) $\downarrow$}} \\
    \cline{3-5}
      & & All & NoErr & Err \\
    \midrule
    ASR & - & 13.63 & \textbf{0.00} & 21.98 \\
    \midrule
    + zero-shot  & Qwen2-7B-I   & 13.66 & 0.46 & 21.74 \\
    \setlength{\parindent}{0.8em} JEA             & Llama-2-7B-H & 16.96 & 5.05 & 24.25 \\
    \midrule
    + finetuned & Qwen2-7B    & \textbf{11.78} & \underline{0.32} & \textbf{18.79} \\
    \setlength{\parindent}{0.8em} JEA      & Qwen3-8B    & 12.09 & 0.37 & 19.26 \\
        & Llama-2-7B    & 12.13 & 0.54 & 19.23 \\
        & Llama-3.1-8B & \underline{11.89} & 0.46 & \underline{18.89} \\
    \bottomrule
  \end{tabularx}
\end{table}

Table~\ref{table:results_by_wer} reports WER on the SAP-Hypo5 test set across multiple experimental settings. The ASR baseline achieves an overall WER of 13.63\% (0\% on the \textit{NoErr} group and 21.98\% on the \textit{Err} group). Applying Qwen2-7B-Instruct for zero-shot correction only yields sight improvements on the \textit{Err} group. On the other hand, Llama2-H trained on HypoParadise—an ASR post-error correction benchmark—fails to outperform the baseline in terms of WER. While zero-shot approaches show clear limitations for reliable WER reduction, supervised fine-tuning (SFT) substantially mitigates these issues: across all models, SFT consistently reduces WER on the \textit{Err} group (from 22\% to 19\%) while maintaining stable performance on the \textit{NoErr} group ($<$0.6\% degradation), which confirms the robustness and transferability across different LLM families for ASR post-correction through domain-specific adaptation.

\begin{table*}
  \centering
  \caption{SAP-Hypo5 test set (\textit{Err} split) results using different Judge-Editor Agents, reported in multiple metrics (\%). Following Table~\ref{table:results_by_wer}, the first row is the ASR baseline. WER: Word Error Rate; Q-Emb: Qwen3-Embedding Similarity; BERT F1: BERTScore F1; MENLI: natural language inference; Intent Acc.: Intent Accuracy; Slot F1: Slot Micro F1}
  \label{table:results_by_multi_metrics}
  \renewcommand{\arraystretch}{1.0}
  \begin{tabularx}{1.0\linewidth}{p{2.25cm}|p{2.1cm}||>{\centering\arraybackslash}X|*{3}{>{\centering\arraybackslash}X}|*{2}{>{\centering\arraybackslash}X}}
    \toprule
    \multirow{3}{*}{\textbf{Method}} & \multirow{3}{*}{\textbf{Agent}}
      & \multicolumn{6}{c}{\textbf{Metrics (\%)}} \\
    \cline{3-8}
      & & \multirow{2}{*}{\textbf{WER $\downarrow$}} &  \multicolumn{3}{c|}{\textbf{Semantic $\uparrow$}} & \multicolumn{2}{c}{\textbf{Task-oriented $\uparrow$}} \\
      \cline{4-8}
      & & & \textbf{Q-Emb} & \textbf{BERT F1} & \textbf{MENLI} & \textbf{Intent Acc.} & \textbf{Slot F1} \\
    \midrule
     ASR  & - & 21.98 & 88.18 & 74.51 & 55.62 & 82.51 & 52.15 \\
    \midrule
    \multirow{2}{*}{\makecell[l]{+ zero-shot JEA}} & Qwen2-7B-I & \textbf{21.74} & 88.22 & 74.65 & 55.90 & 82.64 & 52.70 \\
    & Llama-2-7B-H & 24.25 & \textbf{88.80} & \textbf{75.39} & \textbf{59.90} & \textbf{83.34} & \textbf{53.45} \\
    \midrule
    \multirow{4}{*}{\makecell[l]{+ finetuned JEA} }& Qwen2-7B    & \textbf{18.79} & 89.84 & 77.92 & 62.88 & \textbf{85.45} & 57.85 \\
      & Qwen3-8B    & 19.26 & 89.57 & 77.53 & 62.03 & 84.24 & 57.99 \\
    & Llama-2-7B   & 19.23 & 89.77 & 78.06 & 63.21 & 85.00 & 59.43 \\
     & Llama-3.1-8B & 18.89 & \textbf{89.97} & \textbf{78.35} & \textbf{63.21} & 84.94 & \textbf{59.81} \\
    \bottomrule
  \end{tabularx}
\end{table*}

Table~\ref{table:results_by_multi_metrics} extends the WER analysis with multi-metric results on the \textit{Err} group. In the zero-shot setting, Qwen2-7B-Instruct slightly but consistently outperforms the baseline in all metrics. However, Llama-2-7B-H improves semantic and task-oriented metrics while degrading WER, demonstrating that an improvement in semantic fidelity as measured by five different embedding-based or task-oriented metrics is not necessarily correlated with an improvement in WER.
Fine-tuning on SAP-Hypo5 yields consistent gains across all metrics, while exhibiting model-dependent advantages. Qwen2-7B excels in transcription accuracy with the lowest WER at 18.79\%, whereas Llama3.1-8B delivers superior semantic alignment across all semantic metrics and excels in downstream slot-filling performance. These complementary strengths indicate that different foundation models may specialize in different objectives: some focus on minimizing literal discrepancies against the reference, while others are better at preserving semantic integrity and improving downstream task performance.

\begin{table}[t!]
  \centering
  \caption{Ablation study of Judge and Editor Roles with Qwen2-7B (Instruct variant in the zero-shot setup). Results are reported as WER (\%) on the SAP-Hypo5 test set.}
  \label{table:ablation_on_judge_editor}
  \renewcommand{\arraystretch}{1.0}
  \begin{tabularx}{1.0\linewidth}{p{1.6cm}|*{2}{>{\centering\arraybackslash}X}||*{3}{>{\centering\arraybackslash}X}}
    \toprule
    \multirow{2}{*}{\textbf{Setup}} 
      & \multicolumn{2}{c||}{\textbf{Roles}} 
      & \multicolumn{3}{c}{\textbf{WER (\%) $\downarrow$}} \\
    \cline{2-6}
      & Judge & Editor & All & NoErr & Err \\
    \midrule
    Baseline &  \graycross  & \graycross & 13.63 & 0.00 & 21.98 \\
    \midrule
    Zero-shot & \ding{51} &   \ding{51} & 13.66 & 0.46 & 21.74 \\
    \midrule
    Fine-tuned 
    &  \graycross   & \ding{51} & 13.33 & 0.14 & 21.41 \\
    & \ding{51}  &  \graycross  & 13.25 & \textbf{0.11} & 21.29 \\
               & \ding{51} & \ding{51} & \textbf{11.78} & 0.32 & \textbf{18.79} \\
    \bottomrule
  \end{tabularx}
\end{table}

Further analysis reveals that WER is less robust under domain shifts than semantic or task-oriented metrics, largely due to its sensitivity to non-semantic normalization mismatches. For instance, in Llama-2-7B-H, 37.1\% of samples whose WER increases from zero after post-correction were affected by contraction expansion mismatches. Yet these semantic metrics are not equally sensitive: Q-Emb’s sentence-level design limits its sensitivity to fine-grained errors, while BERTScore and MENLI better reflect subtle semantic variations. Overall, a faithful assessment of post-correction performance requires semantic- and task-level metrics beyond WER. MENLI and Slot F1 exhibit the largest room for improvement, making them particularly suitable for guiding future advances in dysarthric speech recognition and post-error correction.

\subsection{Ablation Study}

To evaluate the role of each component in \textbf{Judge–Editor Agent}, we report an ablation study on Qwen2-7B in Table~\ref{table:ablation_on_judge_editor}. The Judge-only agent selects the best candidate among hypotheses without modification, while the Editor-only agent edits the top-1 hypothesis. Zero-shot Judge–Editor yields little improvement, but fine-tuning greatly boosts performance. The best results arise when Judge and Editor are combined, underscoring their complementary roles in robust dysarthric speech correction.

\vspace*{-6pt}
\section{Conclusion and Future Work}
\label{sec:Conclusion and Future Work}
\vspace*{-4pt}
This paper introduced a Judge–Editor framework that leverages LLMs as agents to refine ASR outputs for dysarthric speech. By combining complementary roles—the Judge preserves reliable spans and the Editor rewrites uncertain ones—a lightweight fine-tuned agent can achieve optimal performance on the SAP-Hypo5 benchmark across multiple metrics. While WER is sensitive to domain shifts, semantic and task-oriented metrics such as MENLI and Slot F1 offer more robust guidance for future progress. Looking ahead, in-context learning, and self-check mechanisms offers promising directions to enhance zero-shot performance and framework robustness. Future work should also explore SAP-specific downstream models and incorporate human-centric evaluation to further verify the clinical and practical utility of the proposed framework.

\section{Acknowledgements}
This research used the Delta advanced computing and data resource which is supported by the National Science Foundation (award OAC 2005572) and the State of Illinois. Delta is a joint effort of the University of Illinois Urbana-Champaign and its National Center for Supercomputing Applications. The SAP dataset was made possible by a grant to the University of Illinois from the AI Accessibility Coalition.

\bibliographystyle{IEEEbib}
\bibliography{strings,refs} 

\end{document}